\documentclass[12pt]{article}

\begin{document}

%********************MAKROS**********************************
\newcommand{\beq}{\begin{equation}}
\newcommand{\eeq}{\end{equation}}
\newcommand{\bea}{\begin{eqnarray}}
\newcommand{\eea}{\end{eqnarray}}

\newcommand{\chii}{\raise.5ex\hbox{$\chi$}}
\newcommand{\R}{I \! \! R}
\newcommand{\N}{I \! \! N}
\newcommand{\C}{I \! \! \! \! C}

\newcommand{\noi}{\noindent}
\newcommand{\vs}{\vspace{5mm}}
\newcommand{\ie}{{\ensuremath{ i.e.\ }}}
\newcommand{\eg}{{\ensuremath{ e.g.\ }}}
\newcommand{\ea}{{\ensuremath{ et~al.\ }}}
\newcommand{\hf}{{\scriptstyle{1 \over 2}}}
\newcommand{\ih}{{\scriptstyle{i \over \hbar}}}
\newcommand{\hi}{{\scriptstyle{ \hbar \over i}}}
\newcommand{\itwoh}{{\scriptstyle{i \over {2\hbar}}}}
\newcommand{\dbrst}{\delta_{BRST}}

% functional derivative
\newcommand{\deder}[1]{{ 
 {\stackrel{\raise.1ex\hbox{$\leftarrow$}}{\delta^r}   } 
\over {   \delta {#1}}  }}
\newcommand{\dedel}[1]{{ 
 {\stackrel{\lower.3ex \hbox{$\rightarrow$}}{\delta^l}   }
 \over {   \delta {#1}}  }}

% partial derivative 
\newcommand{\papar}[1]{{ 
 {\stackrel{\raise.1ex\hbox{$\leftarrow$}}{\partial^r}   } 
\over {   \partial {#1}}  }}
\newcommand{\papal}[1]{{ 
 {\stackrel{\lower.3ex \hbox{$\rightarrow$}}{\partial^l}   }
 \over {   \partial {#1}}  }}

% unspecified derivative
\newcommand{\ddr}[1]{{ 
 {\stackrel{\raise.1ex\hbox{$\leftarrow$}}{\delta^r}   } 
\over {   \delta {#1}}  }}
\newcommand{\ddl}[1]{{ 
 {\stackrel{\lower.3ex \hbox{$\rightarrow$}}{\delta^l}   }
 \over {   \delta {#1}}  }}
\newcommand{\dd}[1]{{  {\delta} \over {\delta {#1}}  }}
\newcommand{\pa}{\partial}

%Remember not to leave any spaces, unless you want flying punctuationmarks.
\newcommand{\eq}[1]{{(\ref{#1})}}
\newcommand{\mb}[1]{{\mbox{${#1}$}}}

%%\newtheorem{theorem}{Theorem}[chapter]
%%\newtheorem{proposition}[theorem]{Proposition}
%%\newtheorem{lemma}[theorem]{Lemma}
%%\newtheorem{claim}[theorem]{Claim}
%%\newtheorem{remark}[theorem]{Remark}
%%\newtheorem{example}[theorem]{Example}

%End of proof box:
\newcommand{\proofbox}{\begin{flushright}
${\,\lower0.9pt\vbox{\hrule \hbox{\vrule
height 0.2 cm \hskip 0.2 cm \vrule height 0.2 cm}\hrule}\,}$
\end{flushright}}
%********************ENDMAKROS**********************************

%\newcommand{\noi}{}
%\newcommand{\vs}{}

\newcommand{\galg}{{\cal G}}
\newcommand{\ggroup}{ G}
\newcommand{\bgalg}{\bar{\cal G}}
\newcommand{\bggroup}{ \bar{G}}
\newcommand{\ci}[1]{\cite{#1}}

%%%%%%%%%%%%%%%%%%%%%%%%%%%%%%%%%%%%%%%%%%%%%%%%%%%%%%%%%%%%%%%%%%%%%%
%%%%%%%%%%%%%%%%%%%%%%%%%%%%%%%%%%%%%%%%%%%%%%%%%%%%%%%%%%%%%%%%%%%%%%
%%%%%%%%%%%%%%%%%%%%%%%%%%%%%%%%%%%%%%%%%%%%%%%%%%%%%%%%%%%%%%%%%%%%%%
%%%%%%%%%%%%%%%%%%%%%%%%%%%%%%%%%%%%%%%%%%%%%%%%%%%%%%%%%%%%%%%%%%%%%%
\thispagestyle{empty}
%\vspace{10cm}
%\begin{center}
\title{\Large{\bf Geometry of the BFV Theorem}}
%\end{center}
%\vspace{4cm}
%\begin{center}
\author{{\sc K.~Bering}\thanks{Address after Sept.\ 1st, 1997: Center for 
Theoretical Physics, MIT, Cambridge MA 02139, USA.}
\\Institute of Theoretical Physics\\
Uppsala University\\P.O. Box 803\\S-751 08 Uppsala\\Sweden\\}
%\end{center}
\maketitle
\begin{abstract}
We describe gauge-fixing at the level of virtual paths in the 
path integral as a non-symplectic BRST-type of flow on the path phase space.
As a consequence a gauge-fixed, non-local symplectic structure  
arises. Restoring of locality is discussed. A pertinent anti-Lie-bracket 
and an infinite dimensional group of gauge fermions are introduced. 
Generalizations to Sp(2)-symmetric BLT-theories are made.
\end{abstract}
%\vspace{1.8cm}
\begin{flushleft}
UUITP-14/97\\hep-th/9709003
\end{flushleft}

\vfill
\newpage

\setcounter{equation}{0}
\section{Introduction}

\vs
\noi
Similarly to the Lagrangian BRST symmetry\ci{refbrst} for 
Yang-Mills theory, Batalin, Fradkin and Vilkovisky
\ci{refbfvpath} developed a BRST formulation of an arbitrary
Hamiltonian gauge theory with reducible first class constraints and
possible open gauge algebra, nowadays known as BFV-BRST 
quantization\ci{refhenneaux}.
The crucial ingredients turn out to be two Grassmann-odd 
objects, the nilpotent BRST charge \mb{\Omega} 
and the gauge fermion \mb{\psi}. The BFV Theorem, which 
is the subject of this article, simply states 
that the partition function \mb{{\cal Z}} is independent 
of the choice of \mb{\psi}.

\vs
\noi
The present work was originally motivated by a concern about the stability
of time locality. It is a well-known fact that time has a special 
status in Hamiltonian theories. One assumes that all virtual paths 
in the path integral are parametrized by the one and 
same global time parameter. As a consequence,
it makes sense to speak of equal-time Poisson brackets. 
In Hamiltonian theories, there is no interaction between two 
different time-slices.  In other words, the symplectic two-form 
\mb{\omega(t)=\omega(z(t),t)} is a {\em function}
of the coordinates taken in the very same time. This is often referred to
as saying that the phase space Poisson structure is ultra-local in time.
So the symplectic structure does not depend on the past 
nor the future.
Nevertheless, in the standard proof\ci{refbfvpath} for the BFV Theorem one
performs an infinitesimal flow generated by a 
bosonic BRST-type vector field 
\beq
 X(t)~ =~ \mu X_ {\Omega(t)}~,\label{brsttypeoftrans}
\eeq
where
\mb{\mu \sim \int_{0}^{T}  \! \!  dt \  \delta_{\rm ext} \psi(t) } 
is a {\em functional}. The flow therefore carries information
about the past and the future of the path. 
In a consistent geometric formulation such flows will violate the
ultra-local ansatz for the two-form \mb{\omega(t)=\omega(z(t),t)}. 
The BRST variation \eq{brsttypeoftrans} is transporting well-defined
ultra-local theories into non-local theories.
Naively, one would expect that the pertinent BRST transformation 
would map ultra-local theories in ultra-local theories.
However, this is not the case.
Our analysis below shows that the non-locality arising in the
canonical measure factor, the Pfaffian, remarkably can be 
recast into an ultra-local form, which turns out to be the 
gauge-fixing term in the action. This solves our original posed question,
and it shows that the formalism is consistent.

\vs
\noi
Besides the time-locally problem mentioned above,
we introduce and give a geometric description of
the gauge fixing flow directly at the level of virtual paths
in the path phase space. 

\vs
\noi
The article is organized as follows: A pertinent anti-Lie-bracket, 
that turns the set of gauge fermions into an infinite-dimensional algebra
is introduced in Section \ref{secalg}. This algebra is interesting in its own 
right from a pure mathematical point of view. 
There we also discuss the 
realizations of the algebra of gauge fermions and its corresponding
group. This paves the way for introduction of a gauge-fixing flow
discussed in Section \ref{secgfflow}. With this in hand we can prove the
BFV theorem in a path space formalism, see Section \ref{secbfv}. 
We have for completeness included a Section \ref{secinf} 
about the infinitesimal gauge-fixing flow.
Finally, we shall in Section \ref{secsp2} generalize the construction to 
$Sp(2)$-symmetric Hamiltonian theories\ci{bltha2}. 
Some of the result in this article 
has been reported on the NATO Advanced Research Workshop, 1997, Zakopane, 
``New Developments in Quantum Field Theory''.

\vs
\setcounter{equation}{0}
\section{Lift to Path Space} 
\label{seclift}

\vs
\noi
The basic philosophy is to consider the collection  \mb{{\cal P}\Gamma} 
of all paths as an infinite-dimensional 
manifold, \ie a geometric object that does not depend 
on the choice of coordinates. For instance, given a volume form,
the partition function is an integral
over this manifold weighted with the Boltzmann factor.
Let us emphasize that both the Boltzmann factor and the volume form 
are scalar objects, that does not depend on the 
choice of coordinates.
In this picture the BRST transformation \eq{brsttypeoftrans} is
a non-Hamiltonian (and non-local) vector field on the manifold 
\mb{{\cal P}\Gamma}, \ie it is not a symplectic or equivalently a 
canonical transformation.
So the symplectic structure is affected 
by the BRST vector field \eq{brsttypeoftrans} and 
it in general becomes non-local.

\vs
\noi
We see that the BRST-type transformation \eq{brsttypeoftrans}
forces us to consider a wider class of theories than the ultra-local ones.
To include non-local theories, we have to lift the
construction from phase space \mb{\Gamma} to a {\em path} phase space
\mb{{\cal P}\Gamma}, thereby giving room for non-trivial Poisson bracket
between different times. In general, non-local theories are senseless,
but we shall see that we can give meaning to a restricted class of
non-local theories.

\vs
\noi
We shall here discard global obstructions and  
assume that objects like the gauge fermion $\psi$ and the 
symplectic potential $\vartheta$ exist globally. So in particular,
our considerations does not include Gribov problems. Also we will
for simplicity consider the case where $\Gamma$ does not have boundaries.
Moreover, we assume that the path integral exists for the studied gauges.

\vs
\noi
Let us introduce a gauge-fixed symplectic two-form 
directly in the path space 
\mb{{\cal P}\Gamma=\prod_{t\in [0,T]} \Gamma}:
\bea
\omega_{\psi}&=&\omega_{\psi}(\gamma)
~=~\hf \int_{\gamma}\int_{\gamma} dt \  dt' \  dz^A(t)\  
\omega_{\psi AB}(t,t') \ \wedge \ dz^B(t') \cr   
&=&\hf\int_{\gamma}\int_{\gamma} dt \ dt'\ \omega_{\psi AB}(t,t') \ 
dz^B(t') \ \wedge \ dz^A(t)  (-1)^{\epsilon_A+1} ~. 
\label{pathspacesymplectic2form}
\eea 
The basic picture is that acting with a BRST type
transformation \eq{brsttypeoftrans} change the symplectic two-form
depending on the gauge-fixing $\psi$.
In other words, \mb{\omega_{\psi}} is a modified
symplectic two-form, whose change is an accumulated effect of
acting with a BRST-type of transformations \eq{brsttypeoftrans}.
We shall derive the explicit formula later. Here we are only
interested in the principle idea.
The inverse \mb{\omega_{\psi}{}^{AB}(t,t')} 
gives rise to a gauge-fixed path space Poisson bracket
\beq
\left\{ F, G \right\}_{\psi} ~=~\left\{ F, G \right\}_{\psi}(\gamma) 
~=~  \int_{\gamma}\int_{\gamma} dt \  dt' \ 
 F \deder{z^A(t)} \omega_{\psi}{}^{AB}(t,t') \dedel{z^B(t')} G~,
\eeq
where \mb{F,G:{\cal P}\Gamma \to \C}.
For generic $\psi$ the elements \mb{\omega_{\psi AB}(t,t')} 
are functionals of the path \mb{\gamma}. 
More precisely, we allow for the following dependence:
A) It could be a {\em functional} of the full path $\gamma$, B)
it could be a function of the two path values $\gamma(t)$ and $\gamma(t')$
and finally C) it could depend explicitly on the times $t$ and $t'$.
In symbols,
\beq
 \omega_{\psi AB}(t,t')
~=~ \omega_{\psi AB}(\gamma,\gamma(t),\gamma(t'),t,t')~.
\eeq
As an obvious boundary condition, when the gauge fermion is turned off,
\mb{\psi=0}, the gauge-fixed symplectic structure should
coincide with the original ultra-local symplectic structure.  That is,
\beq
\omega~=~\omega(\gamma)
~=~\hf \int_{\gamma}\int_{\gamma} dt \  dt' \  dz^A(t)\  
\omega_{AB}(z(t),t) \ \delta(t-t') \ \wedge \ dz^B(t')~.
\eeq

\vs
\noi
In the same manner, instead of looking at classical observables
\mb{f:\Gamma \to \C}, let us consider
classical observables \mb{F:{\cal P}\Gamma \to \C} that are functions on the
{\em path} space \mb{{\cal P}\Gamma}. 
The usual ultra-local  classical observables $F$ are of the form
\beq
 F(\gamma)~=~\int_{\gamma} dt \ f(\gamma(t),t) ~~~~~
{\rm or} ~~~~ F(\gamma)~=~f(\gamma(t_1),t_1) 
\label{clasobsbuildblock}
\eeq
for some function \mb{f: \Gamma \times [0,T] \to \C} and \mb{t_1 \in[0,T]}.
Examples are the BRST-improved Hamiltonian \mb{H_{BRST}(z(t),t)} 
and the BRST charge \mb{\Omega(z(t),t)}. The path space versions read
\bea
 H_{BRST} &=& H_{BRST}(\gamma)
~=~\int_{\gamma} dt \ H_{BRST}(\gamma(t),t)~, \cr 
\Omega&=&\Omega(\gamma)~=~\int_{\gamma} dt\  \Omega(\gamma(t),t)~, \cr 
\{\Omega,\Omega\}&=&0~, 
{}~~~~~~~~~~~~~~~\{\Omega,H_{BRST}\}~=~0 ~.\label{hohoho}
\eea
The classical observables of the form \eq{clasobsbuildblock}
are formally the fundamental building blocks for all classical observables, 
\ie a classical observable $F$ in path space is a formal powerserie 
of the type \eq{clasobsbuildblock}.
As a particular example we allow that the gauge fermion 
\mb{\psi:{\cal P}\Gamma \to \C} is a powerserie of these 
building blocks. 

\vs
\setcounter{equation}{0}
\section{The Space of Gauge Choices}
\label{secalg}

\vs
\noi
Let us analyze the set of gauge fermions in more detail. First of all, 
it turns out that we can define a Grassmann-odd bracket structure among 
the gauge fermions. We have chosen here to start with this construction to 
emphasize its fundamental nature. We shall later see how
this anti-algebra can be realized in the algebra of vector fields as
BRST-like vector fields. When exponentiating to the corresponding groups
we realize the corresponding BRST flow or -- what turn out to be the same 
-- the gauge-fixing flow, see Section \ref{secgfflow}.

\vs
\subsection{An Anti-Lie-Bracket} 

\vs
\noi
An anti-Lie-bracket in the space \mb{C^{\infty}({\cal P}\Gamma)} 
of ``classical'' observables is defined via 
the Poisson bracket and the BRST charge:
\bea
   (F,G) &=& \ih (-1)^{\epsilon_{F}} \{\Omega,F G\}
{} ~=~ \ih  \{F G,\Omega\}(-1)^{\epsilon_{G}+1} \cr
&=& -(-1)^{(\epsilon_{F}+1)(\epsilon_{G}+1)}(G,F)~.
\eea
The anti-Lie-bracket is of odd Grassmann parity and it has ghost number $+1$. 
It satisfies the correct symmetry property, the Jacobi identity,
but not the Poisson property.

\vs
\subsection{Group of Gauge Fermions} 

\vs
\noi
The space of gauge fermions \mb{\psi:{\cal P}\Gamma \to \C}
with  ghost number $-1$ equipped with this anti-Lie-bracket is an
infinite dimensional Lie-algebra, that we denote \mb{\galg}.
The corresponding group \mb{\ggroup} is identified with precisely 
the same space of ghost number $-1$ functions  
\mb{\Psi:{\cal P}\Gamma \to \C}, so \mb{\ggroup=\galg}.
The group \mb{\ggroup} is endowed with an associative product 
\mb{\diamond:\ggroup \times \ggroup \to \ggroup},
\beq
 \Psi_{1} \diamond  \Psi_{2} ~=~   
 \Psi_{1}+ \Psi_{2}+ \ih  \Psi_{1}\{\Omega,\Psi_{2}\}~,
\eeq
and an inversion map \mb{\tau:\ggroup  \to \ggroup},
\beq
\tau(\Psi)~=~-\frac{\Psi}{1+{{i} \over {\hbar}}\{\Omega,\Psi\}}
{}~=~-\frac{\Psi}{1+B}~=~-\Psi \ e^{-b} ~.
\eeq
We have for convenience introduced 
\beq 
b\equiv\ih \{\Omega,\psi\}=\ln(1+B)~,~~~~~~~
B\equiv\ih \{\Omega,\Psi\}=e^{b}-1~.
\eeq
The algebra elements are in general denoted by a lowercase $\psi$ and the 
group element are denoted by uppercase $\Psi$. They are connected via 
the {\em bijective} exponential map \mb{{\rm Exp}:\galg \to \ggroup}:
\beq
  \Psi={\rm Exp}(\psi)=\psi \  e(b)~,~~~~
 \psi={\rm Ln}(\Psi)=\Psi \ l(B)~,
\eeq
where \mb{e} and \mb{l} are
\beq
 e(b)~=~ \int_{0}^{1} d\alpha \
e^{\alpha b}~=~\frac{e^{b}-1}{b}~=~\frac{B}{b}
{}~=~\sum_{n=1}^{\infty}\frac{b^{n-1}}{n!}~,
\eeq
\beq
 l(B)~=~ \frac{\ln(1+B)}{B}~=~ \frac{b}{B}
{}~=~\sum_{n=1}^{\infty}\frac{(-B)^{n-1}}{n}~.
\eeq
The neutral element of the Lie group is \mb{\Psi=0}: 
\beq
  \Psi \diamond 0 = \Psi =0 \diamond \Psi~,~~~~~~~~ 
{\rm Exp}(0)=0~.
\eeq
The exponential map obeys the Baker-Campbell-Hausdorff formula:
\bea
{\rm BCH}(u\psi_{1},u\psi_{2})
&=& {\rm Ln}({\rm Exp}(u\psi_{1})\diamond{\rm Exp}(u\psi_{2})) \cr
&=&u\psi_{1}+u\psi_{1}+\hf(u\psi_{1},u\psi_{2})+{\cal O}(u^3)~, \cr
{\rm Exp}(-\psi)&=&\tau({\rm Exp}(\psi))~.
\eea
This is easiest to prove in an algebra/group representation
-- a so-called realization -- that we describe in the next Section.

%%%%%%%%%%%%%%%%%%%%%%%%%%%%%%%%%%%%%%%%%%%%%%%%%%%%%%%%%%%%%%%%%%%%%%
\vs
\subsection{Group of Diffeomorphisms} 

\vs
\noi
Consider the infinite dimensional Lie group \mb{{\rm Diff}({\cal P}\Gamma)} 
of diffeomorphisms
%\footnote{\underline{Notation}:
%We shall use the word {\em diffeomorphism} 
%as synonymous with an {\em automorphism}, \ie a diffeomorphism 
%that starts and ends on the same manifold. 
%Thus a diffeomorphism in our terminology
%is a purely geometric notion, that does not depend on the local choice 
%of coordinates, despite what the name may perhaps indicate. 
%If we want to change local coordinates, we instead speak of a 
%{\em coordinate transformation}. One may say that a {\em diffeomorphism}
%is an active operation, while a {\em coordinate transformation} is a
%passive operation.} 
\mb{\sigma:{\cal P}\Gamma \to {\cal P}\Gamma}.
The corresponding infinite dimensional Lie algebra 
\mb{{\rm Lie}({\rm Diff}({\cal P}\Gamma))} is identified with the
set of bosonic vector fields 
\mb{X: C^{\infty}({\cal P}\Gamma) \to  C^{\infty}({\cal P}\Gamma)}
with the usual Lie bracket 
\beq
[X,Y][F]~=~X[Y[F]]-Y[X[F]]~.
\eeq
Here we have adapted the common definition \ci{nakahara} of a vector field 
$X$ that it is a linear derivations on the space of functions 
\mb{C^{\infty}({\cal P}\Gamma)}, \ie that it satisfies 
a graded Leibnitz' rule. Although this definition has the correct 
transformation properties under change of coordinate patches
and it works in the general case, it has a complication that
a more elementary definitions of a vector field often doesn't have: 
It relates naturally to the {\em reversed} group 
\mb{({\rm Diff}({\cal P}\Gamma), \stackrel{{\rm op}}{\circ})}
of opposite ordering, where 
\mb{\sigma_{1} \stackrel{{\rm op}}{\circ}\sigma_{2}
=\sigma_{2}\circ\sigma_{1}}. (Or vice-versa, the Lie-algebra of vector 
fields acting from the {\em right} corresponds to the group of usual 
ordering.) To see this, let us write down the pertinent exponential
map \mb{{\rm Exp}} from the algebra 
\mb{({\rm Lie}({\rm Diff}({\cal P}\Gamma)),[\cdot,\cdot])} to the group 
\mb{({\rm Diff}({\cal P}\Gamma), \stackrel{{\rm op}}{\circ})}. 
It is defined as follows. For each vector field 
\mb{X \in {\rm Lie}({\rm Diff}({\cal P}\Gamma))}, there exists a
unique one-parameter family solution \mb{u \mapsto \sigma_{X}(u)}, 
to the first-order differential equation
\beq \left\{
\begin{array}{rcl}
   \frac{d}{du}\sigma_{X}(u)&=&X_{\sigma_{X}(u)} \cr
   \sigma_{X}(u=0)&=&{\rm id}_{{\cal P}\Gamma}
\end{array}
\right.
\eeq
In greater detail, this means that for all path \mb{\gamma} and 
functions \mb{f \in C^{\infty}({\cal P}\Gamma)}
\beq
  \frac{d}{du} f \circ \sigma_{X}(u,\gamma)  
~=~ X_{\sigma_{X}(u,\gamma)}[f]~. 
\label{diffeqinmoredetail}
\eeq
We shall not try to justify in this infinite dimensional problem that for 
a fixed path \mb{\gamma} such solutions \mb{\sigma_{X}(u,\gamma)} exists 
in some interval \mb{u \in [-\epsilon(\gamma),\epsilon(\gamma)]}. 
And even worse, these solutions should exists uniformly in $\gamma$
over path space \mb{{\cal P}\Gamma} for \mb{u \in [0,1]}.
To do this properly, we should give a rigorous definition of
an infinite dimensional manifold \mb{{\cal P}\Gamma} of paths. 
But this would take us to
far away from the main scope of this paper. 
We shall simply assume this can be done. 
Then the exponential map \mb{{\rm Exp}} 
is  defined as \mb{{\rm Exp}(X)\equiv\sigma_{X}(u\!=\!1)}.
One may show that
\bea
\sigma_{X}(u) \circ \sigma_{X}(v)&=&\sigma_{X}(u+v)~,~~~~~~
\sigma_{X}(u)~=~{\rm Exp}(uX)~, \cr
{\rm Exp}(uX)\circ{\rm Exp}(vX)&=&{\rm Exp}((u+v)X)~,~~~~~~
{\rm Exp}(X)^{-1}~=~{\rm Exp}(-X) ~.
\eea
Note that \eq{diffeqinmoredetail} implies that
\beq
   \frac{d}{du}({\rm Exp}(uX)^{*}f)|_{u=0}~=~X[f]~\equiv~{\cal L}_{X}f~.
\eeq
Together with the boundary condition this yields formally
\beq
   f \circ{\rm Exp}(uX) ~\equiv~ {\rm Exp}(uX)^{*}f
 ~=~\sum_{n=0}^{\infty} \frac{1}{n!} (u{\cal L}(X))^{n} f 
 ~\equiv~e^{u{\cal L}(X)}f~.
\eeq
Hence, the formulas for the pull-back and the push-forward read
\bea
  ({\rm Exp}(X))^{*}&=& e^{{\cal L}(X)}~,\cr
 ({\rm Exp}(X))_{*}Y&=&Y \circ \sum_{n=0}^{\infty} \frac{1}{n!} X^{n} 
 ~\equiv~Y \circ e^{\circ X}   ~.
\eea
Moreover, from the reversing property of the pull-back 
\mb{\sigma_{1}^{*} \sigma_{2}^{*}=(\sigma_{2} \circ \sigma_{1})^{*}},
it follows that
\beq
{\rm Exp}(Y) \circ {\rm Exp}(X) 
~=~{\rm Exp}(X) \stackrel{{\rm op}}{\circ} {\rm Exp}(Y)
  ~=~{\rm Exp} ({\rm BCH}(X,Y))~. 
\eeq
This shows that the Lie algebra of vector fields which acts to the left
is naturally associated with the Lie group of {\em opposite} ordering.

\vs
\noi
The exponential map 
\mb{{\rm Exp}:{\rm Lie}({\rm Diff}({\cal P}\Gamma)) 
\to{\rm Diff}({\cal P}\Gamma)}
gives rise to a right action conventionally denoted by a dot
\beq 
 .~:~{\cal P}\Gamma \times {\rm Lie}({\rm Diff}({\cal P}\Gamma)) 
 \to {\cal P}\Gamma~.
\eeq
Its definition and main properties are
\beq
 \gamma.X=({\rm Exp}(X))\gamma~,~~~~~~~ 
\gamma.0=\gamma~,~~~~~~~
 (\gamma.X).Y =\gamma.({\rm BCH}(X,Y))~~.
\eeq

\vs
\noi
In particular, if the path space \mb{{\cal P}\Gamma} is a right
\mb{{\rm Lie}({\rm Diff}({\cal P}\Gamma))} module, \ie a 
\mb{C^{\infty}({\cal P}\Gamma)} vector space
endowed with an action
\beq
[\cdot,\cdot]: {\cal P}\Gamma
\times {\rm Lie}({\rm Diff}({\cal P}\Gamma)) 
 \to {\cal P}\Gamma
\eeq
satisfying 
\bea
[\gamma, F \ X +G \ Y]&=&F(\gamma)[\gamma,X]+G(\gamma)[\gamma,Y]~, \cr
[a\gamma_{1}+b\gamma_{2},X]&=&a[\gamma_{1},X]+b[\gamma_{2},X]~,\cr
[[\gamma,X],Y]-[[\gamma,Y],X]-[\gamma,[X,Y]]&=&0~,
\eea
then the exponential map can be understood by simple
Taylor expansion:
\bea
 ({\rm Exp}(X))\gamma~=~\gamma e^{[~\cdot~,X]}
&=&\sum_{n=0}^{\infty} \frac{1}{n!}
\underbrace{[[\ldots,[\gamma,X]\ldots,X ],X]} ~. \cr
&&~~~~~~~~~~~~n~{\rm commutators}
\eea
However, it is unnecessary to assume this. 

\vs
\setcounter{equation}{0}
\section{The Gauge-Fixing Flow}
\label{secgfflow}

\vs
\noi
We can now construct a Lie algebra homomorphism 
\mb{X:\galg \to {\rm Lie}({\rm Diff}({\cal P}\Gamma))} 
that takes a gauge fermion $\psi$ to a bosonic 
BRST-type vector field \mb{X^{\psi}} by
\beq
X:\psi \mapsto  X^{\psi}\equiv \ih \psi X_{\Omega}~,~~~~~~~~
[X^{\psi_{1}},X^{\psi_{2}}]=X^{(\psi_{1},\psi_{2})}~. 
\eeq
The composition of two such vector fields is remarkably again a 
vector field of the same type
\beq
    X^{\psi_{1}} \circ X^{\psi_{2}}
 ~=~ X^{\psi_{1}\{\psi_{2},\Omega\}}~,
\eeq
because the derivatives of second order cancels.
The corresponding Lie group homomorphism
\beq
\sigma~\equiv~{\rm Exp} \circ X \circ{\rm Ln} ~:~
\ggroup \to {\rm Diff}({\cal P}\Gamma)
\eeq
 reads
\bea
\sigma:~\Psi &\mapsto&\sigma_{\Psi}
 ~=~ {\rm id}_{{\cal P}\Gamma} + \ih  \Psi X_{\Omega}~, \cr
\sigma_{\Psi_{1}\diamond {\Psi_{2}}}
&=&\sigma_{\Psi_{1}}\stackrel{{\rm op}}{\circ}\sigma_{\Psi_{2}}
~=~\sigma_{\Psi_{2}}\circ \sigma_{\Psi_{1}}
~, \cr
\sigma_{\tau(\Psi)}&=&(\sigma_{\Psi})^{-1}~.
\eea
This is called  a realization of \mb{\ggroup}.
As a very important consequence we have a gauge-fixing right group action
\beq
\diamond ~:~ {\cal P}\Gamma  \times \ggroup \to {\cal P}\Gamma
\eeq
directly on the path space:
\bea
  \gamma_{\Psi}~=~  \gamma \diamond\Psi &=& \sigma_{\Psi}(\gamma)
 ~=~\left( {\rm id}_{{\cal P}\Gamma}
 + \ih  \Psi X_{\Omega} \right) \gamma ~,\cr
 \gamma\diamond 0&=&\gamma~, \cr
(\gamma \diamond \Psi_{1}) \diamond \Psi_{2}  
&=&  \gamma \diamond (\Psi_{1} \diamond \Psi_{2}) ~. 
\label{directlyonpath}
\eea
This is the sought-for geometric gauge-fixing. The
gauge fixing can be understood at the level of 
paths as a modification of the individual paths! 

\vs
\noi
Again, the expression 
\mb{\sigma_{\Psi}={\rm id}_{{\cal P}\Gamma} +  \ih \Psi X_{\Omega}} is 
quite symbolic, but it is justified by the formula for
the pull-back:
\beq
(\sigma_{\Psi})^{*}
~=~  ({\rm id}_{{\cal P}\Gamma} + \ih  \Psi X_{\Omega})^{*}
 ~=~1 + \ih  {\cal L}( \Psi X_{\Omega})~. 
\eeq
The expression can be evaluated directly term by term on the manifold if the 
path space \mb{{\cal P}\Gamma} is a module wrt.\ the Lie algebra  
\mb{(C^{\infty}({\cal P}\Gamma),\{\cdot,\cdot\})}.

\vs
\noi
Before we carry on, let us for the sake of completeness also 
define the map 
\bea
\sigma:~\psi &\mapsto&\sigma_{\psi}
 ~=~{\rm Exp}(X^{\psi})
 ~=~{\rm Exp}( \ih \psi X_{\Omega})~:~
\galg ~\to~ {\rm Diff}({\cal P}\Gamma)~, \cr
 \left( \sigma_{\psi} \right)^{-1}&=&  \sigma_{-\psi}~,
\eea
and the corresponding right ``gauge-fixing action'' 
\mb{ {\cal P}\Gamma  \times \galg \to {\cal P}\Gamma}
on the path space \mb{{\cal P}\Gamma} from the Lie 
algebra \mb{\galg}:
\bea
 \gamma.\psi&=&\sigma_{\psi}(\gamma)
 ~=~({\rm Exp}( \ih \psi X_{\Omega}))\gamma~, \cr
\gamma.0&=&\gamma~, \cr
 (\gamma.\psi_{1}).\psi_{2}
&=& \gamma.({\rm BCH}(\psi_{1},\psi_{2}))~.
\eea

\vs
\subsection{Gauge-Fixed Quantities}

\vs
\noi
The gauge-fixed classical observables \mb{F_{\psi}} are defined
via the pull-back of the gauge-fixing map:
\bea
F_{\psi}(\gamma) &=& (\sigma_{\psi}^{*} F) (\gamma)
 ~=~  F (\sigma_{\psi}( \gamma)) \cr
% ~=~F(\gamma+\epsilon) \cr
%&=& \exp[:\int_{\gamma} dt \ 
% \epsilon^{A}(t) \dedel{z^A(t)} :]F(\gamma) \cr
&=& F( \gamma)+ \ih  \Psi(\gamma) \  \{\Omega(\gamma),F( \gamma)\} \cr
&=&  e^{{i \over \hbar} \psi(\gamma)
 \{\Omega(\gamma),\cdot\}}  F( \gamma)~. 
\label{gffct}
\eea
%The normal order symbol means that the exponential function
%should be Taylor expanded with all derivatives to the right.
%\beq
%\epsilon~=~ \sigma_{\psi}(\gamma) - \gamma
%~=~ \Psi(\gamma) \ \{\Omega(\gamma), \gamma\}~,
%\eeq

\vs
\noi
The BRST-improved Hamiltonian \mb{H_{BRST}}
and the BRST charge $\Omega$ are not changed 
by the gauge fixing flow, because of \eq{hohoho}.
The gauge-fixed symplectic two-form and the symplectic  potential 
are remarkably simple, but clearly non-local:
\beq
\omega_{\Psi}~=~\sigma_{\Psi}^{*}\omega
 ~=~\omega + \ih  d\Psi \ \wedge \ d\Omega~,\label{niceone}
\eeq
\beq
\vartheta_{\Psi}~=~\sigma_{\Psi}^{*}\vartheta
 ~=~\vartheta + \ih \Psi  \ d\Omega
 + \ih  d[  \Psi  \ i_{ X_{\Omega}}\vartheta ]~.
\eeq
The gauge-fixed Poisson bracket satisfies
\beq
\{F_{\Psi} , G_{\Psi} \}_{\Psi} 
 ~=~ \left( \{ F, G \} \right)_{\Psi} ~,
\eeq
or  written more explicitly
\bea
\{ F, G \}_{\Psi} &=&\sigma_{\Psi}^{*} 
\left( \{ \sigma_{\tau(\Psi)}^{*}F,
 \sigma_{\tau(\Psi)}^{*}G  \} \right) \cr
&=& \{ F, G \} 
- \ih  \{ F, \Psi \} \frac{1}{1+B} \{\Omega, G \}
- \ih  \{ F,\Omega \} \frac{1}{1+B} \{ \Psi, G \} \cr
&&+ (\ih)^{2}  \{ F,\Omega \} 
 \frac{\{ \Psi, \Psi  \}}{(1+B)^2} \{\Omega, G \} \cr
&=& \{ F, G \} 
- \ih  \{ F, \psi e(b) \} e^{-b} \{\Omega, G \}
- \ih  \{ F,\Omega \} e^{-b} \{ e(b) \psi, G \} \cr
&&+ (\ih)^{2}  \{ F,\Omega \} 
 e^{-b} \{ e(b) \psi, \psi e(b) \} e^{-b} \{\Omega, G \}~.
\label{gfpb}
\eea
These formulas follows by straightforward application of \eq{gffct}. 
The derivation becomes somewhat easier if one first derive \eq{gfpb} for
\mb{G=F} being Grassmann-odd. Then one may use the polarization formula
\mb{\{F,G\}=\hf( \{F+G,F+G\}- \{F,F\}- \{G,G\})} and finally linearity
to extend to functions of both Grassmann-parities. An even simpler
way is the following matrix-calculation. Define, cf.\ \eq{niceone} :
\bea
\omega_{\Psi AB}(t,t') 
&=& \omega_{AB}(t,t')+ \Delta\omega_{AB}(t,t')~, \cr 
 \Delta\omega_{AB}(t,t') &=&
 \ih (\dedel{z^{A}(t)}\Psi) (\Omega \deder{z^{B}(t')})
 + \ih (\dedel{z^{A}(t)}\Omega) (\Psi \deder{z^{B}(t')})~.
\eea
Then the inverse matrix is 
\bea
[\omega_{\Psi}]^{-1}
&=&([1]+[\omega]^{-1}[\Delta\omega])^{-1} [\omega]^{-1} \cr
&=&\sum_{n=0}^{\infty} \left(- [\omega]^{-1}
[\Delta\omega])^{-1}  \right)^{n} [\omega]^{-1}~.
\eea
It is easy to see that this sums up to \eq{gfpb} with \mb{F=z^{A}(t)} and
\mb{G=z^{B}(t')} substituted.

\vs
\noi
The BRST-closed (and therefore BRST-exact) 
quantities are constant under the
BRST gauge-fixing flow. A BRST exact quantity 
\beq
 \{\Omega,F \}~=~\{\Omega, F_{\psi} \}_{\psi} ~,~~~~~~
\Omega_{\psi} ~=~\Omega ~,
\eeq
can be written BRST-exactly wrt.\ the gauge-fixed structure.
As an important consequence the BRST cohomology is not affected by the
gauge-fixing.

\vs
\setcounter{equation}{0}
\section{The BFV Path Integral}
\label{secbfv}

In this Section we shall formally prove the BFV Theorem in
the path space formulation, by inspecting the various parts of
the path integral. We have already argued that 
the BRST improved Hamiltonian \mb{H_{BRST}} 
is not modified by the gauge fixing flow. Below we shall
analyze the measure part and the kinetic part.

\vs
\subsection{The Volume Form}

\vs
\noi
Let us denote the volume form by
\mb{ \Omega_{\rm vol} ={\cal D}z \ {\rm Pf}(\omega_{\cdot\cdot})}.
The gauge-fixed volume form can remarkably be written as 
\beq
   \sigma^{*}_{\psi} \Omega_{\rm vol}
 ~=~ \Omega_{\rm vol}  \ \frac{1}{1+B}
 ~=~\Omega_{\rm vol} \ e^{-{{i} \over {\hbar}}\{\Omega,\psi\}}~.
\label{crucialvolform}
\eeq
It is perhaps surprising that the gauge-fixing term,
which can be view as a way to remove zero-directions in the Hessian
of the action,
from a geometric point of view has little to do with the action. 
In fact, the above shows that it emerges from the measure part, 
completely dictated by the gauge-fixing flow. 
Because of the importance of \eq{crucialvolform},
let us indicate a proof in the case of a finite number $2n$ 
of purely bosonic variables \mb{z^{A}(t)=z^{I}}, 
where \mb{(A,t)=I=1,\ldots , 2n}, so that
\beq
  \Omega_{\rm vol}= \frac{1}{n!}~ \omega^{\wedge~n} 
 ~=~ \frac{1}{n!} ~\omega \wedge \ldots \wedge \omega~.
\eeq
If there are fermionic variables present, the volume form is no longer 
the $n$'{th} exterior power of the symplectic two-form,
and the analysis becomes more involved. However, let's stick to
the bosonic case. The gauge-fixed top-form is, cf.\ \eq{niceone}, 
\bea  
\sigma^{*}_{\Psi} \Omega_{\rm vol}
&=&  \frac{1}{n!} ~\sigma^{*}_{\Psi}\omega \wedge 
\ldots \wedge\sigma^{*}_{\Psi} \omega \cr
&=& \sum_{r=0}^{n} \frac{1}{(n-r)!r!}~  \omega^{\wedge~(n-r)}  \wedge
\left( \ih  d\Psi \ \wedge \ d\Omega\right)^{\wedge~ r} \cr
&=&  \sum_{r=0}^{n} \frac{1}{(n-r)!r!}~ \omega^{\wedge~(n-r)} ~
 \wedge  \sum_{I_{1}, \ldots, I_{r}}~
( \ih dz^{I_{1}} \frac{\delta\Psi}{\delta z^{I_{1}}} ~ \wedge ~ d\Omega )~
  \wedge \cr
&&~~~~~~~~~~~~~~~~~~~~~~~~~~~  \ldots \wedge~
( \ih dz^{I_{r}}  \frac{\delta\Psi}{\delta z^{I_{r}}} ~ \wedge ~ d\Omega )~.
\eea
Because of covariance, one may resort to Darboux coordinates
\mb{z^{I}=(q^{i},p_{i})}, \mb{i=1,\ldots , n}. Then
\beq
 \{q^{i},p_{j} \} ~=~ \delta^{i}_{j} ~,~~~~~~~~~~~~~~~~~~~~~~~~~~~~ 
\omega~=~ dp_{i} \wedge dq^{i}~.
\eeq
A simple combinatorial argument yields that
\bea   
  \sigma^{*}_{\Psi} \Omega_{\rm vol}
&=&  (dp_{1}~ \wedge~ dq^{1})~ \wedge
 \ldots \wedge ~(dp_{n}~ \wedge~ dq^{n}) \cr
&& ~~~~~~~~~\times
 \sum_{r=0}^{n}  \sum_{i_{1}, \ldots, i_{r}}
 (-\ih) ( \frac{\delta\Psi}{\delta q^{i_{1}}}~ 
\frac{\delta\Omega}{\delta p_{i_{1}}}
 -  \frac{\delta\Psi}{\delta p_{i_{1}}}~ 
\frac{\delta\Omega}{\delta q^{i_{1}}} ) 
\times  \cr 
&& ~~~~~~~~~ \ldots\times (- \ih) ( \frac{\delta\Psi}{\delta q^{i_{r}}}~ 
\frac{\delta\Omega}{\delta p_{i_{r}}}
 -  \frac{\delta\Psi}{\delta p_{i_{r}}}~ 
\frac{\delta\Omega}{\delta q^{i_{r}}} ) \cr
&=&  \Omega_{\rm vol} \sum_{r=0}^{n} (-B)^{r}
 ~=~ \Omega_{\rm vol} \frac{1-(-B)^{n+1}}{1+B}~.
\eea
Now we formally take the \mb{n=\infty} limit to derive \eq{crucialvolform}.

\vs
\subsection{The Kinetic Part}

\vs
\noi
We have already analyzed the variation of the Hamiltonian part and
the measure under the gauge-fixing flow. Let us now draw our attention 
to the last piece of the path integral, namely the kinetic term.
It is handy to introduce a vector field $T$ that calculates the difference
between the total time derivative and the explicit time derivative: 
\beq
T~\equiv~ {{d} \over {dt}}-{{\partial} \over {\partial t}}
 ~=~\int_{\gamma} dt \  \dot{\gamma}^{A}(t) \dedel{z^{A}(t)}
 ~~:~~ C^{\infty}({\cal P}\Gamma) \to C^{\infty}({\cal P}\Gamma)~.
\eeq
The kinetic term can be written as the contraction of the
symplectic potential with this vector field: 
\beq
 i_{T} \vartheta(\gamma)= 
\int_{\gamma} dt \  \vartheta_{A}(t) \dot{\gamma}^{A}(t)~,~~~~~~
 \vartheta(\gamma)=\int_{\gamma} dt \ \vartheta_{A}(t) dz^{A}(t)~.
\eeq
We can now write the gauge-fixed BFV path integral as
\beq
{\cal Z}_{\psi}~=~\int \Omega_{\rm vol}~ 
e^{{i \over \hbar} (i_{T} \vartheta -H_{BRST}-\{\Omega,\psi\})}~.
\eeq
We would like to only sum over paths in the path integral so that
the kinetic term for these paths is left invariant under the 
gauge-fixing flow. 
In fact, if this is fulfilled, we have realized gauge-fixing as 
an internal diffeomorphism of the path integral.\footnote{
Perhaps it is helpful a this point to recall that every integral
\mb{I=\int f(x) \ dx=\int \! f(\sigma(x))  \ d\sigma(x)}
by the very definition is diffeomorphism invariant. 
\mb{x} is just a dummy variable.}
Thus the path integral cannot depend on the gauge fermion and we
can conclude the BFV Theorem. To achieve this conclusion,
one should restrict the path space by only allowing
for paths satisfying the following BRST boundary condition
\beq
 \left. (i_{X_{\Omega}} \vartheta)(\gamma) \right|_{t=0}
+ \left. \Omega (\gamma) \right|_{t=0}
 ~=~   \left. (i_{X_{\Omega}} \vartheta)(\gamma) \right|_{t=T}
+ \left. \Omega (\gamma) \right|_{t=T}~.
\label{kinetictermbc}
\eeq
Because of the formula
\beq
 i_{X_{\Omega}} \vartheta (\gamma)
 ~=~\int_{\gamma} dt \  \vartheta_{A}(t)  X_{\Omega}^{A}(t) 
\eeq
and because \mb{\Omega} is BRST-invariant, we have
\bea
 X_{\Omega}[ i_{X_{\Omega}} \vartheta ] (\gamma)
&=& \int_{\gamma} \int_{\gamma} dt \ dt' \  
X_{\Omega}^{B}(t') \dedel{z^B(t')} 
[   \vartheta_{A}(t) \ X_{\Omega}^{A}(t)  ] \cr
&=&\hf \int_{\gamma} dt \  
[X_{\Omega}, X_{\Omega}]^{A}(t) \ \vartheta_{A}(t) (-1)^{\epsilon_{A}} \cr
&&+ \int_{\gamma} \int_{\gamma} dt \ dt' \  
X_{\Omega}^{A}(t) \ X_{\Omega}^{B} (t') \
  \dedel{z^B(t')} \vartheta_{A}(t) \cr
&=& \hf \int_{\gamma} dt \  
X_{ \{ \Omega, \Omega \} }^{A}(t)  \ \vartheta_{A}(t) (-1)^{\epsilon_{A}} \cr
&&+ \hf\int_{\gamma}\int_{\gamma} dt \ dt' \  
 X_{\Omega}^{A}(t) \  X_{\Omega}^{B}(t') \
\omega_{BA}(t',t) ~=~0~. 
\eea
Together with \mb{ X_{\Omega}[\Omega]=0} this yields that 
the above boundary condition on the paths is stabile 
under the gauge-fixing action of \mb{\galg}.
We will assume that a vector field
\beq
X~:~\gamma \mapsto X_{\gamma}
~=~\int_{\gamma} dt  \ X^{A}(t)   \dedel{z^A(t)}  
\eeq
has the following dependence on the path \mb{\gamma}:
\beq
 X^{A}(t)~=~  X^{A}(\gamma,\gamma(t),t)~.
\eeq
Under the assumption that the original symplectic potential 
\mb{\vartheta_{A}(t)=\vartheta_{A}(\gamma(t))}
is ultra-local and that it has no explicit time dependence
(and therefore the symplectic two-form is also of that kind), one
derives that a generic vector field $X$ changes the kinetic term 
with
\beq
X[i_{T} \vartheta] (\gamma)
 ~=~ \left. i_{X} \vartheta\right|_{\gamma,\gamma(t=T),t=T}
 - \left. i_{X} \vartheta\right|_{\gamma,\gamma(t=0),t=0}
+ \int_{\gamma} dt \   X^{A}(t) \  \omega_{AB}(t) \  \dot{z}^{B}(t)  ~.
\eeq
Assuming that \mb{\Omega} is ultra-local and that it has 
no explicit time dependence
and using the boundary condition \eq{kinetictermbc},
we therefore conclude that the kinetic term is invariant under
the gauge-fixing flow \mb{X=\psi X_{\Omega}}.

%\vs
%\noi
%The gauge-fixing flow \mb{\sigma_{\psi}:{\cal P}\Gamma \to{\cal P}\Gamma} 
%on the paths is defined as
%\beq
%\gamma_{\psi} ~=~  \sigma_{\psi}(\gamma)
%~=~e^{\psi(\gamma) \{\Omega(\gamma),\cdot\}} \gamma
%~=~ \gamma + \epsilon
%~=~\exp[:\int_{\gamma} dt \
%\epsilon^{A}(t) \dedel{z^A(t)} :]\gamma ~.
%\eeq
%where 
%\beq
%\epsilon~=~ \sigma_{\psi}(\gamma) - \gamma
%~=~\int_{0}^{1} d\alpha \ 
%{\cal L}\left( e^{\alpha b}\psi \  X_{\Omega} \right)\gamma
%~=~ \psi(\gamma) \ e(b(\gamma)) \ 
%\{\Omega(\gamma), \gamma\}~,
%\eeq
%Note that (er korrekt) 
%\beq
%\{\Omega(\gamma), \gamma_{\psi} \} 
%~=~ e^{b(\gamma)} \{\Omega(\gamma), \gamma\}~.
%\eeq 

\vs
\subsection{Infinitesimal Gauge-Fixing Flow}
\label{secinf}

\vs
\noi
Above we have argued from a formal point of view
that turning on or turning off gauge-fixing does not
change the value of the path integral. In real applications,
like quantum field theory, turning off gauge-fixing makes the
path integral ill-defined, as a result of integrating over an 
infinite gauge-volume. 
That is why gauge-fixing is introduced in the first place.
It is therefore desirable to have an infinitesimal analysis.

\vs
\noi
Consider the map 
\beq
L_{\rm Exp} ~:~ \galg \to {\rm Aut}(\ggroup) ~:~
\psi \mapsto L_{\rm Exp(\psi)}~,
\eeq
where \mb{L_{\Psi}} is the left multiplication map with a fixed \mb{\Psi}
\beq
L_{\Psi}(\Psi')~=~\Psi \diamond \Psi'~.
\eeq
%%In a similar manner consider the realization
%%\beq
%% \sigma  ~:~ \galg \to {\rm Diff}({\cal P}\Gamma)  ~:~
%%\psi \mapsto \sigma_{\psi}~.
%%\eeq
Then by definition
\beq
L_{\rm Exp(0)}~=~{\rm id}_{\ggroup}
%%%~,~~~~   \sigma_{0}~=~{\rm id}_{{\cal P}\Gamma}~.
\eeq
Let us consider an infinitesimal change 
\mb{\psi \to \psi + \delta\psi}
in the gauge fermion.
The relevant object is the differential of
\mb{L_{\rm Exp}} in \mb{\psi=0},
\beq
(L_{\rm Exp})_{*0} ~:~
 T_{0}\galg \to T_{{\rm id}_{\ggroup}} {\rm Aut}(\ggroup)~. 
\eeq
It reads
\beq
\Psi + \delta\Psi~=~\left((L_{\rm Exp})_{*0}~\delta\psi\right) (\Psi) 
 ~=~h_{\Psi}(\delta\psi) \diamond \Psi~.
\eeq
Here the function \mb{h : \galg \times T_{0}\galg \to  \galg}
is 
\bea
h_{\psi} (\delta\psi) &=& (\Psi + \delta\Psi) \diamond \tau(\Psi) 
 ~=~\frac{\delta\Psi}{1+B} 
 ~=~\frac{ {\rm Exp}_{*\psi}(\delta\psi)}{1+B} \cr
&=& e^{-b} e(b) \ \delta\psi 
+  e^{-b} e'(b) \ \psi \ \{\Omega,\delta\psi\} \cr
&=& \frac{1-e^{-b}}{b} \delta\psi
 + \frac{b-1+e^{-b}}{b^2} \psi \ \{\Omega,\delta\psi\}~.
\eea
We have identified the tangent space 
\mb{T_{{\rm id}_{\ggroup}} {\rm Aut}(\ggroup) }
with maps from the group \mb{\ggroup} that to each element $\Psi$ of
the group \mb{\ggroup} give an element in \mb{\galg \diamond \Psi}.

\vs
\noi
So the infinitesimal change in the gauge fixing flow 
\mb{\sigma_{\psi + \delta\psi} \circ \sigma_{-\psi} } is
governed by a BRST-type vector field, cf.\ \eq{directlyonpath},
\beq
 X^{h_{\psi}(\delta\psi)} ~=~
 h_{\psi}(\delta\psi) X_{\Omega}~. 
\eeq
Note that
\beq
\{\Omega, h_{\psi} (\mu) \} =\{\Omega,\mu \} ~,~~~~~~
h_{\alpha\psi}(\psi) =\alpha \ \psi~.
\eeq
%\beq (er korrekt)
%e(-b) ~=~ e^{-b} e(b) ~.
%\eeq 
We can now derive the infinitesimal change of various objects
under an infinitesimal change \mb{\psi \to \psi + \delta\psi} 
of the gauge fermion caused by the vector field 
\mb{X^{ {h}_{\psi}(\delta\psi) }}. For instance, 
the measure changes according to 
\beq
{\rm div}_{\rho} X^{h_{\psi}(\delta\psi)}
 ~=~- \ih \{\Omega,\delta\psi\}~.
\eeq
This is precisely the infinitesimal change in the gauge-fixed action.

\vs
\setcounter{equation}{0}
\section{The Sp(2) Case}
\label{secsp2}

\vs
\noi
The whole construction can easily be extended to the $Sp(2)$ case\ci{bltha2}.
We shall thus be brief and just state the construction.

\vs
\noi
In the $Sp(2)$ case we have two odd BRST 
charges \mb{ \Omega^{a}}, \mb{a=1,2}, 
that are mutual orthogonal in the Poisson bracket sense: 
\beq
  \{ \Omega^{a}, \Omega^{b} \}~=~0~.
\eeq
We define a Grassmann-even bracket
\beq
[F,G]~=~  F \ k(G) -  k(F) \ G 
 ~=~-(-1)^{\epsilon_{F}\epsilon_{G}}[G,F]~,
\eeq
where $k$ is a Grassmann-even nilpotent second order operator
\bea
 k&=& \itwoh \epsilon_{ab}X_{\Omega^{a}} \circ X_{\Omega^{b}}~,
 ~~~~~k^2~=~0~, \cr
 k(F)&=&\itwoh \epsilon_{ab} \{ \Omega^{a}, \{ \Omega^{b},F \} \}
 ~=~ \itwoh   \{ \{ F, \Omega^{a} \}, \Omega^{b} \} \epsilon_{ab}~.
\eea
The Grassmann-even bracket satisfies the usual symmetry property, 
the Jacobi identity, but not the Poisson property.
We define a degenerate metric structure (a covariant $2$-tensor)
\bea
g(F,G)&=&   k(F) \ G +  F \ k(G)-  k(FG) ~=~
\ih \{ F , \Omega^{a} \} \epsilon_{ab}  \{ \Omega^{b}, G \} \cr
&=&+(-1)^{\epsilon_{F}\epsilon_{G}} g(G,F)~.
\eea
$k$ and $g$ satisfy the following relations
\beq
 g(k(F),G) ~=~0~,~~~~~~~
k(g(F,G)) ~=~ 2 \ k(F) \ k(G)~,
\eeq
\beq
 g(F,g(G,H))~=~\hf g(F,G) \ k(H) 
+\hf g(F,H) \ k(G)(-1)^{\epsilon_{G}\epsilon_{H}}~.
\eeq
We have a Jacobi-like identity:
\beq
 g(F,g(G,H))-(-1)^{\epsilon_{F}\epsilon_{G}} g(G,g(F,H))
 ~=~g([F,G],H)~.
\eeq
%\beq
% g(F,g(G,H))+(-1)^{\epsilon_{F}\epsilon_{G}} g(G,g(F,H))
%~=~-2g(F,G) \ k(H) -  k(F) \ g(G,H)
% - (-1)^{\epsilon_{F}\epsilon_{G}} k(G) \ g(F,H)
%\eeq
%\beq
% wrong g(F,g(F,G)) ~=~g(F,F) \ k(G)    \ g(\phi,F)
%\eeq
The gauge is now chosen by specifying a gauge {\em boson}\footnote{
Our gauge boson is {\em minus a half} of the BLT gauge boson.} $\phi$.
The even bracket \mb{[\cdot,\cdot]} is a Lie-bracket in the
algebra \mb{\bgalg} of gauge bosons.
The space \mb{\bggroup=\bgalg} of gauge bosons becomes a Lie group
by introducing a product and an inversion
\bea
 \Phi_{1} \diamond  \Phi_{2} &=&   
 \Phi_{1}+ \Phi_{2} + \Phi_{1} \ k(\Phi_{2})~,\cr
\tau(\Phi)&=&-\frac{\Phi}{1+\bar{B}}~=~-\Phi \ e^{-\bar{b}} ~,
\eea
where 
\bea 
\bar{b}&=&\itwoh \epsilon_{ab} \{ \Omega^{a}, \{ \Omega^{b},\phi \} \}
 ~=~k(\phi)~=~\ln(1+\bar{B})~, \cr
\bar{B}&=&\itwoh \epsilon_{ab} \{ \Omega^{a}, \{ \Omega^{b},\Phi \} \}
 ~=~k(\Phi)~=~e^{\bar{b}}-1~,
\eea
and \mb{\phi} and \mb{\Phi} are connected via 
the bijective exponential map 
\mb{{\rm Exp}:\bgalg \to \bggroup}
\beq
  \Phi={\rm Exp}(\phi)=\phi \  e(\bar{b})~,~~~~~~~~
 \phi={\rm Ln}(\Phi)=\Phi \ l(\bar{B})~.
\eeq
We can define a homomorphism 
\mb{\Psi_{a}:\bgalg \to \galg_{a}}
that transforms a gauge boson \mb{\phi} into 
a gauge fermion\footnote{To be precise, 
\mb{\Psi_{a}(\phi)=\hf\psi} is 
{\em half} the corresponding
gauge fermion of the standard BFV-BRST construction.}
 \mb{\Psi_{a}(\phi)} of the kind \mb{a=1,2}
(corresponding to the generator \mb{\Omega^{a}})
\bea
    \Psi_{a}(\cdot) &=& \epsilon_{ab} X_{\Omega^{b}}(\cdot) \cr
  \Psi_{a}(F)&=&\epsilon_{ab} \{ \Omega^{b} , F \}
     ~=~ (-1)^{\epsilon_{F}}\{ F , \Omega^{b} \} \epsilon_{ba} ~,\cr
   \Psi_{a}\left([\phi_{1},\phi_{2}]\right)
  &=&( \Psi_{a}(\phi_{1}), \Psi_{a}(\phi_{2}))^{a}~, \cr
  \Psi_{a}\left(\Phi_{1} \diamond \Phi_{2}\right)
  &=& \Psi_{a}(\Phi_{1}) \diamond  \Psi_{a}(\Phi_{2})
\eea
(no sum over $a$.). Note that 
\beq
 \ih  \{\Omega^{a} , \Psi_{b}(F) \}~=~\delta^{a}_{b} \ k(F)~.
\eeq
The algebra homomorphism 
\mb{\bar{X}:\bgalg \to {\rm Lie}({\rm Diff}({\cal P}\Gamma))}
is the sum of the two BRST homomorphisms 
\beq
\bar{X}:\phi \mapsto  \bar{X}^{\phi}\equiv g(\phi, \cdot)
=\ih \Psi_{a}(\phi) X_{\Omega^{a}} ~,~~~~~~~~
[\bar{X}^{\phi_{1}},\bar{X}^{\phi_{2}}]=\bar{X}^{[\phi_{1},\phi_{2}]}~. 
\eeq
The corresponding Lie group homomorphism
\mb{\bar{\sigma}\equiv{\rm Exp} \circ \bar{X} \circ{\rm Ln} :
\bggroup \to {\rm Diff}({\cal P}\Gamma)} reads
\bea
\bar{\sigma}:~\Phi &\mapsto&\bar{\sigma}_{\Phi}
 ~=~ {\rm id}_{{\cal P}\Gamma} +   g( \Phi,\cdot )
 + \hf g(\Phi,\Phi) \ k(\cdot ) ~, \cr
\bar{\sigma}_{\Phi_{1}\diamond {\Phi_{2}}}
&=&\bar{\sigma}_{\Phi_{1}} \stackrel{{\rm op}}{\circ}
 \bar{\sigma}_{\Phi_{2}}~, \cr
\bar{\sigma}_{\tau(\Phi)}&=&(\bar{\sigma}_{\Phi})^{-1}~.
\eea
The pull-back  map \mb{ (\bar{\sigma}_{\Phi})^{*}} is
\beq
(\bar{\sigma}_{\Phi})^{*} ~=~
1 + \ih \Psi_{a}(\Phi) \ {\cal L}(  X_{\Omega^{a}})
+ \itwoh\epsilon_{ab} \  g(\Phi,\Phi) \ {\cal L}(  X_{\Omega^{a}})
 \circ {\cal L}(  X_{\Omega^{b}})~. 
\eeq
The gauge-fixed symplectic two-form reads
\beq
\omega_{\Phi}~=~\bar{\sigma}_{\Phi}^{*}\omega
 ~=~\omega + \ih  d\Psi_{a}(\Phi) \ \wedge \ d\Omega^{a}~.
\eeq
The gauge-fixed volume form reads
\beq
   \bar{\sigma}^{*}_{\phi} \Omega_{\rm vol}
 ~=~  \Omega_{\rm vol} \ \frac{1}{(1+\bar{B})^{2}}
 ~=~  \Omega_{\rm vol}  \ e^{-{{i} \over {\hbar}} 
\epsilon_{ab} \{ \Omega^{a}, \{ \Omega^{b},\phi \} \}}~.
\eeq

%%%%%%%%%%%%%%%%%%%%%%%%%%%%%%%%%%%%%%%%%%%%%%%%%%%%%%%%%%%%%%%%%%%%%%
%%%%%%%%%%%%%%%%%%%%%%%%%%%%%%%%%%%%%%%%%%%%%%%%%%%%%%%%%%%%%%%%%%%%%%
%%%%%%%%%%%%%%%%%%%%%%%%%%%%%%%%%%%%%%%%%%%%%%%%%%%%%%%%%%%%%%%%%%%%%%
%%%%%%%%%%%%%%%%%%%%%%%%%%%%%%%%%%%%%%%%%%%%%%%%%%%%%%%%%%%%%%%%%%%%%%

\vspace{1cm}
\noindent
{\sc Acknowledgements:}~
The Author would like to thank I.A.~Batalin, P.H.~Damgaard, A.M.~Semikhatov,
A.P.~Nersessian and O.M.~Khudaverdian for fruitful discussions.
The work is partially supported by Nordita.

\end{document}